\documentclass[11pt,twoside]{article}
\usepackage{asp2004}\usepackage{epsf}\usepackage{epsfig}
\begin{document}

\title{Physical conditions in the Homunculus}

\author{Gary J. Ferland \& Nick Abel}
\affil{Physics, University of Kentucky}

\author{Kris Davidson}
\affil{Astronomy, University of Minnesota}

\author{Nathan Smith}
\affil{CASA, University of Colorado}

\begin{abstract}
Conditions within the Homunculus nebula around Eta Car are determined
by many of the same physical processes that occur in molecular clouds
in the interstellar medium. But there is one major exception -- we
know when the ejection occurred and something about its composition
and initial state. The gas was warm, ionized, and dust-free when it
was located within the star's atmosphere and it is currently cold,
molecular, and dusty. It undertook this transformation in a bit over
150 years. It offers a laboratory for the study of physical processes
in a well-constrained environment. We derive a photoionization model
of the Homunculus nebula that reproduces many of its observed
properties. We conclude by outlining how observations of the
Homunculus could address basic problems in the physics of the
interstellar medium.
\end{abstract}

\section{Introduction}

Eta Carinae, one of the most luminous stars in the Galaxy, offers a
laboratory in which a variety of physical phenomena can be
studied. The star has undergone at least one episode of substantial
mass loss, and is likely to end its life as a supernova. Understanding
the physics that occurs within the ejecta will offer insight into the
initial stages of chemical enrichment of the interstellar medium. Both
molecules and grains are known to have formed in the Homunculus
nebula, gas that was ejected in the nineteenth century. The physical
conditions in this nebula are the subject of this paper, along with a
sketch of how Eta Car might be used as a test bed to understand
critical interstellar processes.

\section{An illustrative model of the Homunculus}

The Homunculus nebula is seen in the optical by reflected starlight
and is one of the brightest objects in the sky in the mid-IR, showing
that substantial amounts of dust are present (Westphal {\&} Neugebauer
1969). In the infrared, emission lines of [Fe~{\sc ii}] and H$_2$ are
seen and have been traced across the nebula in long-slit observations
(Smith 2002). These show a double-shell structure, with Fe$^{+}$
present in the inner parts of the Homunculus, and H$_2$ in the outer
zone. Additionally, the Homunculus shows a double-shell structure in
the dust color temperature distribution, with a cool outer shell at
$\sim$140 K and a warmer inner shell at $\sim$200-250 K (Smith et
al. 2003). These observations help constrain any model of the
envelope.

We model the shell as a constant-density layer with an inner radius of
1.7$\times$10$^{17}$ cm and a thickness of 10$^{17}$ cm, appropriate
for material in the wall of the southeast polar lobe along our line of
sight to the star (Smith 2002; Smith et al.\ 2003). The electron
density in the inner Fe$^{+}$ region, deduced from infrared
[Fe{\sc~ii}] line ratios, is $\sim$10$^4$ cm$^{-3}$ (Smith 2002). The
hydrogen density must be substantially higher here since the gas is
mainly neutral. From the geometry of the nebula and the presumed total
mass of $\sim$12~M$_{\odot}$ for the entire Homunculus (Smith et al.\
2003), we infer a hydrogen density of $\sim$10$^{6}$ cm$^{-3}$. Of
course, this depends on the gas-to-dust mass ratio, taken to be 100
(but see below). For these parameters the column density through the
shell is $N_H$=10$^{23}$ cm$^{-2}$.

The dust that is clearly present in the nebula will be the catalyst in
forming the observed H$_2$. The measured visual extinction is $A_V
\approx $ 4 mag, (e.g., Davidson \& Humphreys 1997), corresponding to
an extinction per unit column density of $A_{V}$/$N_{H}\approx 4
\times 10^{-23}$ mag cm$^{2}$. The extinction is observed to be grey
and the dust color temperature in the shell is near the equilibrium
blackbody temperature (Whitelock et al. 1983; Hillier et al. 2000;
Smith et al. 1998, 2003), suggesting that the grains are large. We
assume that the grains are similar to those seen in Orion's Veil,
$R$=5, and assume that only the silicate component is present. This
large $R$ is produced by a grain size distribution that is lacking in
small grains, which will affect the H$_2$ formation rate and grain
photoelectric heating.

The ratio of C/O is less than unity in the ejecta, (Davidson et
al. 1986), suggesting that the chemistry will be dominated by
oxygen-bearing species once formation of CO is complete, and that the
chemistry will eventually lead to oxygen-rich solids, motivating our
use of the silicate grain type.  The observations of silicate features
in the infrared (e.g., Gehrz et al.\ 1973), and the absence of a
graphite feature in the ultraviolet (Viotti et al.\ 1989) supports
this idea. For simplicity we leave the silicate dust to gas ratio at
its ISM value, which corresponds to an extended source
$A_{V}$/$N_{H}\approx9\times10^{-23}$ mag cm$^{2}$. The grain size
distribution and dust to gas ratio will affect the details of our
calculations, as well as clumping of the material, but not the overall
results.

The assumed gas-phase abundances are listed in Table 1, along with
other parameters. Relative to H, He is overabundant, while O is highly
underabundant, presumably due to partial CNO cycling. The N/H ratio,
roughly ten times solar, corresponds to the conversion of nearly all C
and O into N (Davidson et al.\ 1986; Smith \& Morse 2004). Most C is
expected to be in the form of the $^{13}$C isotope.

We assume that the stellar continuum is represented by an interpolated
20,000~K CoStar atmosphere with a total luminosity of 5$\times$10$^{6}
L_{\odot}$. We add a high-energy component corresponding to a
3$\times$10$^{6}$ K blackbody with a luminosity of 30 $L_{\odot}$
(e.g., Corcoran et al. 2001). The lack of a prominent H~{\sc ii}
region shows that few hydrogen-ionizing photons strike the inner edge
of the nebula, most likely due to photoelectric absorption by the
stellar wind. We extinguish the net continuum by photoelectric
absorption due to a neutral layer of 10$^{21}$ cm$^{-2}$ to account
for this. Some high-energy photons are transmitted and they help drive
the chemistry. The incident stellar continuum is shown in Figure 1.

We also include the galactic background cosmic ray ionization
rate. The actual ionization rate may be higher if radiative nuclei are
present. Cosmic rays have effects that are similar to X-rays -- they
provide ionization that helps drive the chemistry.

\section{Calculations}

We simulate the conditions within the nebula using the development
version of Cloudy, last described by Ferland et al. (1998).

Recent updates to Cloudy include an improved molecular network that
allow for calculations deep in molecular clouds. Some of these
improvements are discussed in Abel et al.\ (2004). The developmental
version of Cloudy currently predicts molecular abundances for $\sim$70
molecules involving H, He, C, O, N, Si, and S. Approximately 1000
reactions are in the network, with most reaction rates taken from the
latest version of the UMIST database. Our predictions are in good
agreement with other codes that are designed to predict conditions in
PDRs.

The computed ionization, thermal, and molecular structure are shown in 
Figures 2 and 3. The emitted continuum is the solid line in Figure 1.

Hydrogen is predominantly atomic at the illuminated face of the
cloud. We assume that no H-ionizing radiation escapes from the stellar
wind, but the H$^+$ density at the illuminated face is quite sensitive
to the transfer of the incident stellar continuum in the Lyman
lines. If it is bright in these lines then hydrogen can become ionized
by a two-step process. An excited state is populated by absorption of
a Lyman line, which then decays into the H$^0$ 2$s$ level. The Balmer
continuum can photoionize atoms in this state, creating a thin region
with H$^+$ and a high electron density.  The Lyman lines quickly
become self-shielding and the process is no longer important, although
a small amount of H$^+$ is produced across the nebula by cosmic ray
and X-ray ionization.

H$_2$ forms at depth of $\sim$3.5$\times$10$^{16}$ cm, where the
Lyman-Werner bands become optically thick, the continuum between
L$\alpha$ and the Lyman limit is heavily extinguished, and the
destruction rate of H$_2$ goes down dramatically. As Figure 1 shows,
little light escapes at short wavelengths.  Grains are the dominant
opacity across the cloud, helping shield H$_2$ and allowing for
efficient formation by catalysis on grain surfaces. The Fe$^+$ profile
is also shown, indicating an anti-correlation between Fe$^+$ and
H$_2$. Observations (Smith 2002) show that Fe$^+$ and H$_2$ are
segregated, occupying the inner and outer zones of the Homunculus
walls, respectively, which generally agrees with the structure in
Figure 2.

The formation of large amounts of H$_2$ initiates the formation of
heavy-element molecules (see Figure 2 and Table 2). H$_2$ is a step in
the formation of H$_{2}^+$ and H$_{3}^+$, the highly reactive
ion-molecules that undergo ion-neutral reactions to form molecules
containing heavier elements. Large amounts of CO form when its
electronic bands become self-shielding. For this calculation, CO fully
forms at depth of $\sim$9$\times$10$^{16}$ cm. We assume a C/O
abundance ratio of 0.5. At the shielded face the
$n$(CO)/$n$(C$_{tot})$ ratio is nearly unity and a significant amount
of O is in the form of OH.

Nitrogen is strongly enhanced in the ejecta, and several nitrogen-bearing 
molecules are shown in Table 2 and Figure 2. As expected, N$_{2}$ is the 
dominant molecule, although only a small amount of N is in this form -- N 
remains predominantly atomic. 

The gas temperature is shown in Figure 3. It lies in the range
$\sim$50~K to $\sim$300~K and is typical of a PDR. The temperature is
mainly maintained by a balance between grain photoionization heating
and cooling by fine structure lines of C, O, Si, and Fe. The
temperature falls at the point where H$_{2}$ forms due to the strong
absorption by its electronic transitions and also by atomic C. In the
coldest regions heating by cosmic rays becomes important, together
with cooling by CO rotation transitions.

The Homunculus appears ``lumpy'', showing that the envelope does not
have constant density or pressure. These calculations show two
possible sources of local instabilities which might help form
blobs. The radiative acceleration, mainly caused by the absorption of
the incident continuum by grains and lines, exceeds 3$\times$10$^{-3}$
cm s$^{-2}$ at the illuminated face but falls to 3$\times$10$^{-6}$ cm
s$^{-2}$ at the shielded face. This suggests that significant
radiative acceleration can occur over the $\sim$10$^2$ yr lifetime of
the ejecta. It may be Rayleigh-Taylor unstable because of the
decreasing acceleration. The thermal balance is a second source of
instability -- the temperature derivative of the net cooling, defined
as cooling minus heating, is negative across much of the ejecta; this
material is thermally unstable. Both instabilities will be the focus
of future work.

The emitted spectrum is shown as the solid line in Figure 1. Thermal
infrared emission from grains is prominent, along with CO lines in the
mm.  The 10 $\mu$m silicate feature is present since the calculation
only included a silicate component. A detailed comparison with
observations would help constrain the model further, and so help
deduce properties such as the dust abundance and composition.

\section{The Future}
\label{sec:mylabel1}

Our main purpose is to point to directions for future work on the
Homunculus, with the goal of using it as a laboratory to understand
basic physical processes. The Homunculus is an especially well-posed
problem. The ejecta were once part of a hot stellar atmosphere and so
must have initially been warm, ionized, and dust-free. Today it is
cold, dusty, and molecular.  How did it go between these states during
the time since its ejection?

This dust could not have formed in the atmosphere of Eta Car itself,
since the energy density temperature is above the condensation
temperature of most solids. Thus the dust now seen in reflection is
most likely to have formed in the material after being expelled from
the star. The cycle of dust destruction and formation is still poorly
understood (Draine 1990). These newly formed dust grains have a large
size, and measurements of the extinction curve across the spectrum
would help quantify their radii. A comparison between intensities of
the thermal IR continuum and H$_{2}$ or CO lines would measure the
dust-to-gas ratio. Infrared spectral features can also reveal the dust
composition. This is especially important in light of recent work
suggesting that supernovae are important sources of new grains in the
galaxy (Morgan et al. 2003).

The Homunculus is predicted to be predominantly molecular. A molecular
inventory could be obtained from UV observations of electronic
absorption lines or from the many prominent rotational emission lines
that are expected in the IR -- mm. We have a good idea of the initial
gas-phase chemical composition so this inventory would test current
chemical reaction networks, and especially the theory of H$_{2}$
formation on grain surfaces. The chemistry will be affected by
additional cosmic ray ionization if radioactive nuclei are present in
the ejecta, and also by the isotopic variations -- most C should be
$^{13}$C rather than $^{12}$C. Can the chemistry test these
assumptions?

Finally, the prominent structures seen in the Homunculus can test
dynamical theories. Could thermal or radiation driving instabilities
play a role?

Acknowledgments: Research into the physical processes of the ISM is
supported by NSF (AST 0307720) and NASA (NAG5-12020).  N.S.\ was
supported by NASA through grant HF-01166.01A from STScI, which is
operated by AURA, Inc., under NASA contract NAS 5-26555.

\newpage 

TABLE 1 -- Model parameters
\begin{table}
\begin{tabular}
{|p{80pt}|p{92pt}|p{82pt}|p{82pt}|}
\hline
 Parameter &Value &Parameter &  \\
(Homunculus) & &(Stellar) & \\
\hline
He/H& 
0.40& 
Wind extinction& 
$N_{H}$=10$^{21}$ cm$^{ - 2}$ \\
\hline
C/H& 
4.0$\times $10$^{ - 5}$& 
Atmosphere& 
Interp.\ Costar \\
\hline
N/H& 
1.5$\times $10$^{ - 3}$& 
$T_{eff}$ & 
20,000K \\
\hline
O/H& 
7.6$\times $10$^{ - 5}$& 
$L$& 
5$\times $10$^{6}$ L$_{\odot}$ \\
\hline
Remaining & 
Solar& 
Corona& 
 \\
elements & & & 
 \\
\hline
$n_{H}$ & 
10$^{6}$ cm$^{ - 3}$& 
Atmosphere& 
blackbody \\
\hline
Inner radius& 
1.7$\times $10$^{17}$ cm& 
$T_{eff}$ & 
1.3$\times $10$^{6}$ K \\
\hline
Thickness& 
10$^{17}$ cm& 
$L$& 
30 L$_{\odot}$ \\
\hline
Filling factor& 
1& 
& 
 \\
\hline
Covering factor& 
0.03& 
& 
 \\
\hline
Grains& 
R=5 Orion silicate& 
& 
 \\
\hline
Cosmic ray& 
Galactic & 
& \\
&Background & &  \\
\hline
\end{tabular}
\label{tab1}
\end{table}

TABLE 2 -- Predicted column densities (cm$^{ - 2})$
\begin{table}[htbp]
\begin{tabular}
{|p{49pt}|p{45pt}|p{50pt}|p{45pt}|p{53pt}|p{45pt}|}
\hline
Species& 
Log $N$& 
Species& 
Log $N$& 
Species& 
Log $N$ \\
\hline
H$^{0}$& 
22.53& 
N$_{2}$& 
14.64& 
H3$^{ + }$& 
12.55 \\
\hline
H$_{2}$& 
22.52& 
CH$_{2}$& 
13.98& 
HCO$^{ + }$& 
12.50 \\
\hline
N$^{0}$& 
20.18& 
CH& 
13.45& 
NO& 
12.48 \\
\hline
O$^{0}$& 
18.86& 
CH3$^{ + }$& 
13.27& 
SiO& 
12.42 \\
\hline
H$^{ + }$& 
18.50& 
HCS$^{ + }$& 
13.09& 
HCN& 
12.41 \\
\hline
C$^{ + }$& 
18.27& 
CS& 
13.06& 
HeH$^{ + }$& 
12.23 \\
\hline
C$^{0}$& 
18.24& 
CN& 
12.88& 
SiN& 
12.21 \\
\hline
Fe$^{0}$& 
18.17& 
CH$_{3}$& 
12.87& 
OH& 
12.18 \\
\hline
Fe$^{ + }$& 
18.13& 
CH$_{4}$& 
12.80& 
NS$^{ + }$& 
12.12 \\
\hline
CO& 
17.54& 
H$_{2}$O& 
12.63& 
CH$_{5}^{ + }$& 
12.02 \\
\hline
\end{tabular}
\label{tab2}
\end{table}

\begin{figure}[!ht]
\plotone{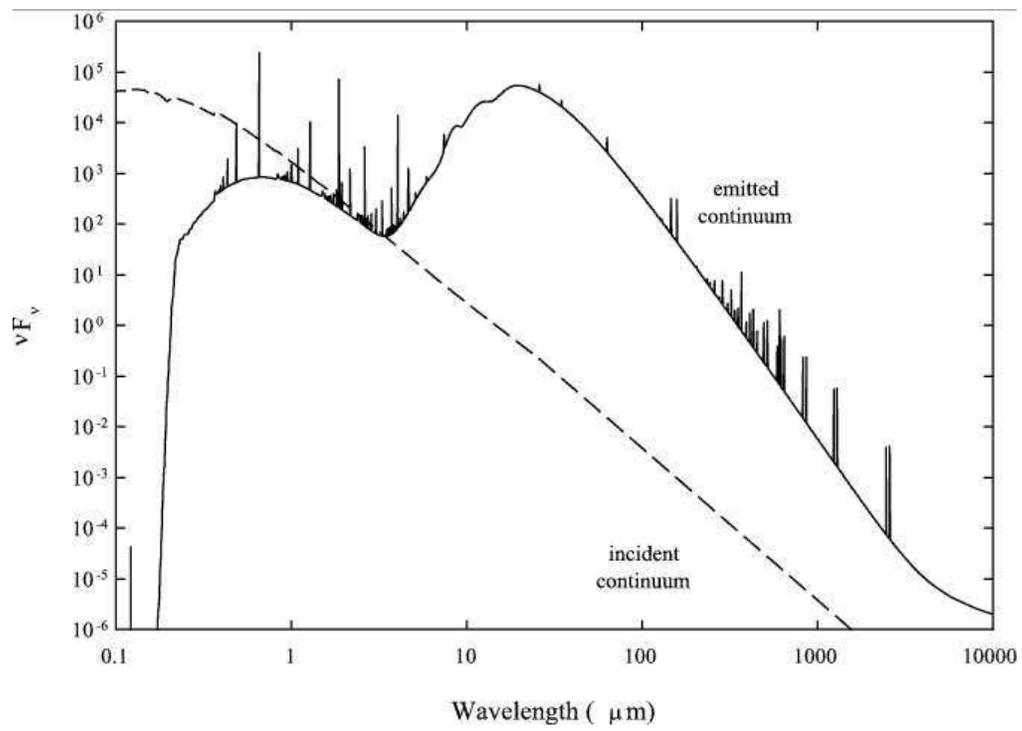}\label{fig1}
\caption{The incident stellar continuum, emitted by the central star,
is shown as a dashed line. The predicted emergent continuum is the
solid line.}
\end{figure}

\begin{figure}
\plotone{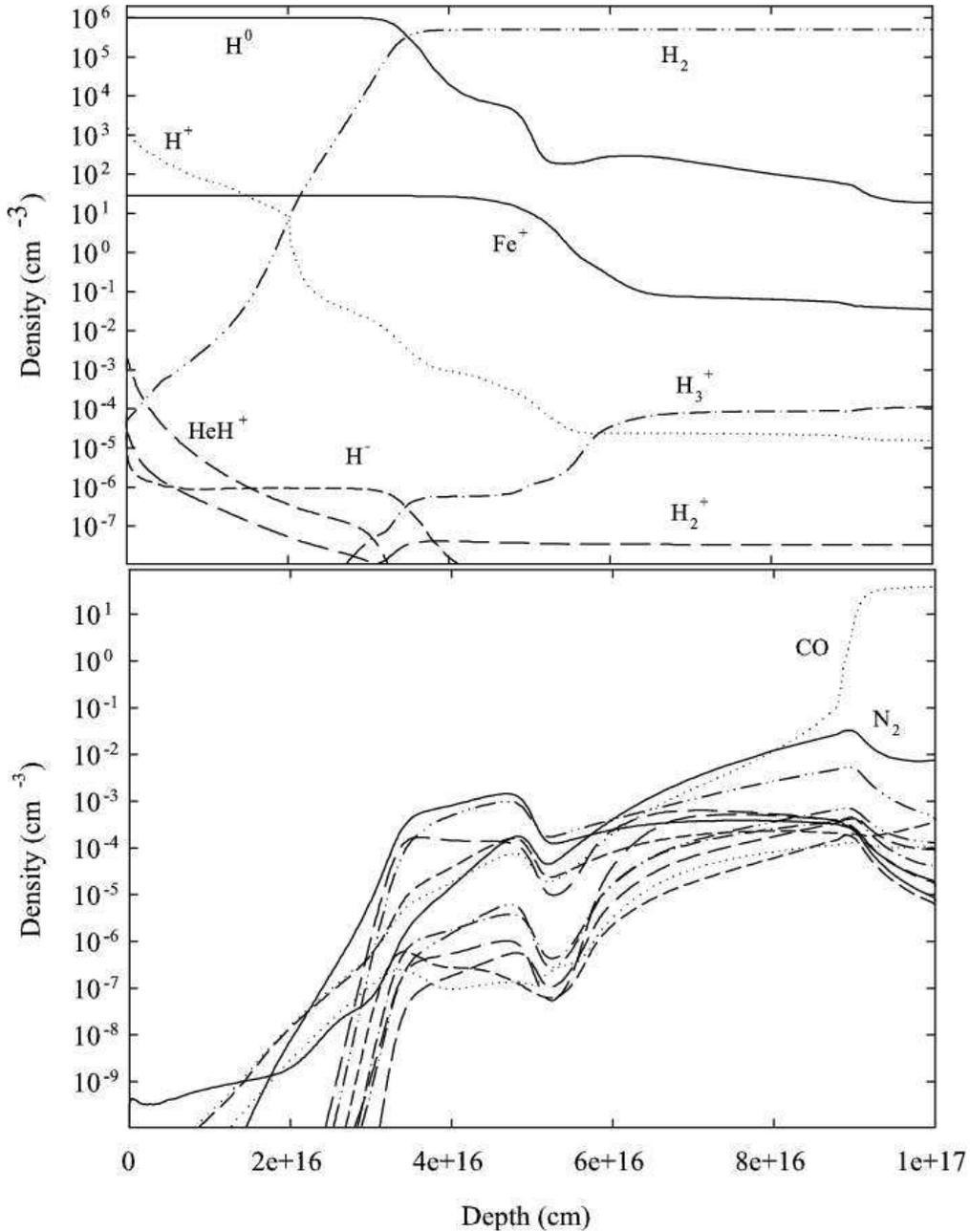}\label{fig2}
\caption{The predicted molecular, atomic, and ionic densities vs depth
into the shell. The upper panel shows that the inner half of the shell
emits Fe~II while H$_{2}$ is found in the outer two-thirds, in
agreement with observations. The lower panel gives densities of the
major molecular constituents. In the outer-most region CO is the
dominant molecule, while N$_{2}$ is the dominant H-carrying
species. The text identifies the remaining molecules. At the shielded
face the molecules, in order of decreasing abundance, are CO, N$_{2}$,
CH$_{2}$, H$_{2}$O, SiO, CS, NO, H$_{3}^{+}$, CH$^{4}$, SiN ,
HS$^{+}$, and CH$_{5}^{+}$.}
\end{figure}

\begin{figure} 
\plotone{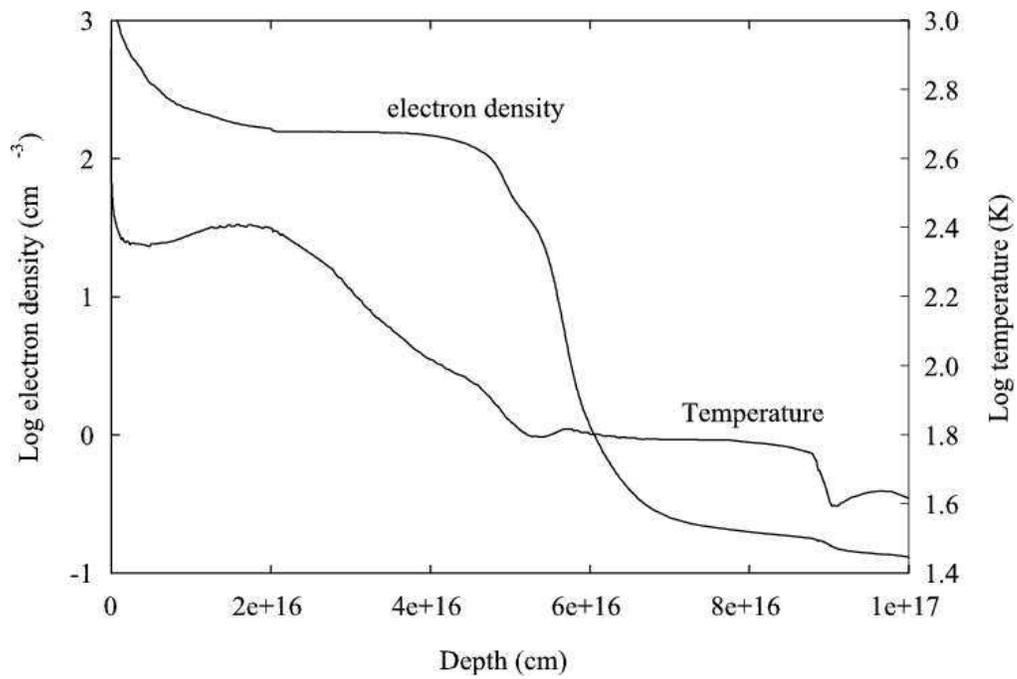}\label{fig3}
\caption{Figure 3, The predicted electron temperature and density
structure. The total hydrogen density was assumed to be constant.}
\end{figure}

\end{document}